\documentstyle[seceq,wrapfig,epsf]{ptptex}

\def\Journal#1#2#3#4{{#1} {\bf #2} (#4), #3}
 
\def\PTP{\em Prog. Theor. Phys.}
\def\NCA{\em Nuovo Cim.}

\def\NPB{{\em Nucl. Phys.} {\bf B}}
\def\PLB{{\em Phys. Lett.}  {\bf B}}
\def\PRL{\em Phys. Rev. Lett.}
\def\PRD{{\em Phys. Rev.} {\bf D}}
\def\ZPC{{\em Z. Phys.} {\bf C}}

\def\be{\begin{equation}}
\def\ee{\end{equation}}
\def\bea{\begin{eqnarray}}
\def\eea{\end{eqnarray}}

\title{%
An Analysis of $\pi\pi$-Scattering Phase Shift 
\\
and
Existence of ${\bf \sigma(555)}$ particle
}

\author{%
Shin {\sc Ishida},
Muneyuki {\sc Ishida}$^{*}$,
Hiroyuki {\sc Takahashi}
\\
Taku {\sc Ishida}$^{**,****}$,
Kunio {\sc Takamatsu}$^{***}$, and
Tsuneaki {\sc Tsuru}$^{****}$
}

\inst{%
Atomic Energy Research Institite, College of Science and Technology
\\
Nihon University, Tokyo 101
\\
$^*$Department of Physics, University of Tokyo, Tokyo 113
\\
$^{**}$Institute for Cosmic Ray Research, University of Tokyo
Tanashi 188
\\
$^{***}$
Department of Engineering, Miyazaki University, Miyazaki 988-12
\\
$^{****}$
National Laboratory for High Energy Physics (KEK), Tsukuba 305
}

\recdate{%
November 22, 1995
}

\abst{%
In most of the Nambu$\cdot$Jona-Lasinio(NJL)-type models, 
realizing the hidden chiral symmetry, the existence of 
a scalar particle $\sigma$
is needed with
a mass $m_\sigma=2 m_q$, as a partner of
the Nambu-Goldstone boson $\pi$. 
However, the results of many analyses on $\pi\pi$ phase-shift
thus far made
have been negative for its existence.\\
\hspace*{\parindent}
In this paper we re-analyze the phase-shift, 
applying a new method, the interfering amplitude method,
which treats the ${\cal T}$-matrix directly and describes multi-resonances 
in conformity with the unitarity. As a result, 
the existence of $\sigma$ has been strongly suggested
from the behavior of the $\pi \pi \rightarrow \pi \pi$ phase shift
between the $\pi\pi$- and the $K\overline{K}$- thresholds, 
with mass = $553.3\pm 0.5_{st} {\rm MeV}$
and width= $242.6 \pm 1.2_{st} {\rm MeV}$. 
The most crucial point in our analysis
is the introduction
of a negative background phase, possibly reflecting a ``repulsive core" in
$\pi\pi$ interactions.\\
\hspace*{\parindent}
The properties of $f_0(980)$ are also investigated from 
data including those over the $K\overline{K}$ threshold.
Its mass is obtained as $993.2 \pm 6.5_{st} \pm 6.9_{sys} {\rm MeV}$.
Its width is about a hundred MeV, although 
this depends largely on the treatment of the
elasticity and the $\pi \pi \rightarrow K\overline{K}$ phase shift, 
both of which may have large experimental uncertainties.
}

\begin{document}

\maketitle

\section{Introduction}
Chiral symmetry has a long history in particle physics
and has played a central role in understanding the
spectroscopy of hadrons with light flavors: In particular
it is now widely accepted that the $\pi$-meson (or pseudo-scalar
meson nonet) has the properties of a Nambu-Goldstone boson,
corresponding to the dynamical breaking of chiral symmetry
(D$\chi$SB) in the massless limit of QCD.

In the Nambu$\cdot$Jona-Lasinio (NJL) model\cite{rf:NJL} and its
extended version (ENJL)\cite{rf:enjl}
\footnote{
In the ENJL model is predicted
the existence of NG boson, scalar meson,
and moreover, vector and axial-vector meson nonets.}
adapted to the quark model, which simply realizes
the physical situation of D$\chi$SB, the existence
of the $\sigma$-meson (or scalar meson nonet) is predicted 
as a chiral partner of
$\pi$ (or a pseudo-scalar meson nonet), 
with iso-spin I=0 and mass=$2m_q$ ($m_q$ being
the constituent quark mass).

It is also well-known that, 
in the investigation of nuclear forces
with one-boson-exchange potential,\cite{rf:obe}
an iso-singlet scalar particle
with mass of about 500MeV, which can
possibly be identified with the $\sigma$
(or a strongly correlated two-pion state \cite{rf:fichi}), 
is necessary.
However, the existence of $\sigma$ 
as a resonant particle has not yet been generally 
accepted.
A major reason for this is due to
the analyses of $\pi\pi$ phase-shift
obtained from the
high-statistics data of a CERN-M\"unich experiment\cite{rf:CM}
in 1974,
in which the phase shift $\delta_0^0$ up to $m_{\pi\pi}=1300{\rm MeV}$ turned
out to be only 270$^\circ$.
Accordingly, after subtracting a rapid contribution of the resonance 
$f_0(980)$$(180^\circ)$,
there remains only $90^\circ$, which is 
insufficient for $\sigma$ around $m$=$2m_q$=$500\sim600{\rm MeV}$, and
many analyses thus far made on the phase shift
have yielded conclusions against the existence of $\sigma$.
\cite{rf:ff,rf:morg,rf:aniso}
\footnote{However, see the analyses\cite{rf:beve,rf:zou,rf:torn}
which suggest the existence of $\sigma$.}

Reflecting this situation, 
the non-linear realization of chiral symmetry 
now seems to be widely
believed rather than the simple linear representation
in the NJL model.
In this case, the existence of $\sigma$ (or a scalar
nonet in ENJL) is unnecessary,
\footnote{See, {\em e.g.}, Refs. \citen{rf:NLS} and
\citen{rf:skir}.}
which is represented as a function
of $\pi$ (or a pseudo-scalar meson nonet).

On the other hand, the possible existence of $\sigma$ has been
suggested from various viewpoints \cite{rf:kunih3}
both theoretically and phenomenologically. 
\cite{rf:menne,rf:svec,rf:shim,rf:kamin}
In particular the importance of $\sigma$ 
in relation with the D$\chi$SB has been
argued extensively by Refs.\citen{rf:sca,rf:kunih0,rf:klim}.

Here, it may be worthwhile to note that
in a recent 
$pp$ central collision experiment,
an event concentration in the I=0, $S$-wave $\pi\pi$ channel
is seen \cite{rf:ti} in the region of $m_{\pi\pi}$ around 
$500\sim 600{\rm MeV}$, which is too large to be explained as
a simple ``background"
and seems strongly to suggest the existence of $\sigma$
\footnote{Owing to Watson's final state interaction theorem,
the $\pi\pi$ 
production amplitude of all types of reactions is closely related to
the amplitude of $\pi\pi$ scattering. Accordingly, if this concentration
is actually due to the $\sigma$-particle production, the
corresponding phase shift
ought to be observed in $\pi\pi$ scattering.}
(a brief review is given in \S 4(B)). 

Considering all the above situations, 
a more rigorous
re-analysis of $\pi\pi$-phase shift
seems to be necessary.

In this paper, the authors 
will perform this by applying a new method of analysis, 
the Interfering-Amplitude (IA) method, which is able to treat partial-wave
multi-resonances directly in the ${\cal T}$-matrix (instead of in the
most conventionally used ${\cal K}$-matrix) in conformity with unitarity in 
the final state $\pi\pi$-interactions. As a result,
evidence for the existence of the $\sigma$-meson is shown
from the characteristic behavior of the
$\pi\pi\rightarrow\pi\pi$ phase shift under the
$K\overline{K}$ threshold.  
A keystone for this is the introduction of a negative
background phase, due to an unknown repulsive force
between pions, which might suggest the existence of a repulsive core
in the $\pi\pi$ interaction. 
Here it is notable that the existence of a similar negative background phase
has been established phenomenologically in the case of the $N$-$N$-interaction.

In \S 2, 
the IA Method is introduced.
Its relation to other approaches, such as 
the ${\cal K}$-matrix method, is discussed in \S 4(A).

In \S 3, 
the $\pi\pi$ scattering phase shift is re-analyzed by applying the IA method. 
It is especially
shown that the characteristic features of phase shifts 
from the $\pi\pi$- to the $K\overline{K}$- thresholds,
including a small bump around 750MeV, are reproduced well with
our new background phase and the $\sigma$ with mass of 555MeV.

The properties of $f_0(980)$ are also examined with the data
including those over the $K\overline{K}$ threshold
in this section.
Its width attains a value from tens MeV\cite{rf:morg}
to hundreds of MeV,\cite{rf:zou} 
largely depending upon the treatment of the
elasticity and the $\pi \pi \rightarrow K\overline{K}$ phase shift, 
both of which have large experimental uncertainties.

In \S 4(C) we give some arguments on the repulsive core in the 
$\pi\pi$-interactions, a possible physical
origin of the negative-phase background. It is pointed out that
our best fit value of the core radius is 
almost equal to the structural size of the pion, 
showing a similarity to the case of the $N$-$N$ interaction, where the core
radius has an order of the size of a nucleon.

In \S 4(D) the existence of new types of
$0^{++}$ and $1^{++}$ nonets is suggested. 
These are outside of the 
normal L-excited meson nonets.
$\sigma$ seems to have properties such that it should be assigned 
as a member of this new type of scalar nonet.

\section{Partial-wave multi-resonance and interfering-amplitude method}
In extracting resonances due to physical particles from scattering 
amplitudes, we usually fit partial wave amplitudes   
with the Breit-Wigner(BW) form.
The scattering matrix $S$ must satisfy
unitarity, which means conservation of probability
from initial to final states:
\begin{eqnarray}
SS^{\dag} = S^{\dag} S =1.
\label{eq:unis}
\end{eqnarray}
We shall 
treat the $\pi\pi$ $S$-wave scattering
in case of two relevant channels:
$\pi\pi$ and $K\overline{K}$, denoted as channels 1 and 2,
respectively. In this case the $\rho$ matrix has a diagonal form
\begin{eqnarray}
\rho&=&diag(
\frac{|{\bf p}_1|}{8\pi\sqrt{s}}, 
\frac{|{\bf p}_2|}{8\pi\sqrt{s}}),\nonumber \\ 
|{\bf p}_1 |&=&\sqrt{\frac{s}{4} -m_\pi^2},\  
|{\bf p}_2 |=\sqrt{\frac{s}{4} -m_K^2},
\label{eq:rhom}
\end{eqnarray}
The partial wave scattering matrix $S^l$ is defined as
\begin{eqnarray}
S_{\bf ji}&=& (2\pi)^4 \delta^{(4)}(P_j-P_i)\sqrt{\frac{1}{\rho_j\rho_i}}
              \sum_{l=0}^{\infty}\frac{2l+1}{2} P_l
({\bf n} \cdot {\bf n}') S_{ji}^l,
\label{eq:defsl}
\end{eqnarray}
where the suffix {\em j(i)} represents the final (initial) channel,
 ${\bf n}'$(${\bf n}$) is a unit vector of
3-momentum of the first particle in the final (initial) system.
The bold suffices {\bf j}, {\bf i} include momenta 
of the system.

The unitarity equation (\ref{eq:unis})
is represented, in terms of $S^l$, as
\begin{eqnarray}
S^l_{jk}S^{l\dag}_{ki} =\delta_{ji};\ \ \ l=0,1,2,... .
\label{eq:unisl}
\end{eqnarray}
The $S^l$ matrix is symmetric due to time reversal invariance
\begin{eqnarray}
S^l_{ji}=S^l_{ij}.
\label{eq:tri}
\end{eqnarray}
The ${\cal T}$-matrix and its partial wave component
${\cal T}^l$ are usually defined, for relevant
systems of two spinless particles as
\begin{subeqnarray}
S_{\bf ji}&=& \delta_{\bf ji}-i(2\pi)^4 \delta^{(4)}(P_j-P_i){\cal T}_{\bf ji},
\\
{\cal T}_{\bf ji}&=& \sum_{l}(2l+1)P_l
({\bf n} \cdot {\bf n}') {\cal T}_{ji}^l.
\label{eq:st}
\end{subeqnarray}
The $S^l$ is related to ${\cal T}^l$ as
\begin{eqnarray}
S_{ji}^l &=& \delta_{ji} - 2i\sqrt{\rho_j\rho_i}{\cal T}_{ji}^l.
\label{eq:pwuni}
\end{eqnarray}
Equations (\ref{eq:unisl}) and (\ref{eq:tri}) are equivalent, respectively, 
to the following conditions
on ${\cal T}_{ji}^l$:
\begin{eqnarray}
i({\cal T}_{ji}^l&-&{\cal T}_{ji}^{l\dag} )
=2({\cal T}^l \rho {\cal T}^{l\dag})_{ji},
\label{eq:unit}\\
{\cal T}^l_{ji} &=&{\cal T}^l_{ij}.
\label{eq:trit}
\end{eqnarray}
In the case with only one resonance, 
these conditions are satisfied by the BW amplitude.
However, in the case of ``multi-resonance scattering"
(when many resonances exist with the same quantum number),
if we represent the partial wave 
${\cal T}$ matrix as a simple sum of corresponding 
BW formulae, they are not satisfied.

Here we give a new method for describing the scattering amplitude 
in conformity with the unitarity
relation in this case.

We shall treat directly the partial wave amplitude 
defined as
\begin{eqnarray}
a_{ji}^l &\equiv& -\sqrt{\rho_j \rho_i}{\cal T}_{ji}^l,
\label{eq:defa}
\end{eqnarray}
which is related to $S^l$ as
\begin{eqnarray}
S_{ji}^l &=& \delta_{ji} + 2ia_{ji}^l.
\label{eq:auni}
\end{eqnarray}

In the following we treat only $a^l_{ji}$ (or $S^l_{ji}$) with $l=0$, 
and the superscript $l$ is omitted.

\subsection{one-channel multi-resonance scattering amplitude}
First let us consider the one channel case. 

A basic idea of our method is that a phase shift $\delta (s)$ of the
scattering amplitude is essentially only due to physical particle resonances 
through the BW formulae, and the phases of multi-resonances
contribute additively to $\delta (s)$. 
Thus it is represented as
\begin{eqnarray}
\delta (s)= \delta^{Re} =
\stackrel{(1)}{\delta }+
\stackrel{(2)}{\delta }+
\stackrel{(3)}{\delta }+\cdots\ ,
\label{eq:delta}
\end{eqnarray}
where
$\stackrel{(i)}{\delta }$ is a
real phase coming from the i-th resonance.

In the case with one-resonance, the scattering matrix is given as
\begin{eqnarray}
\stackrel{(1)}{S} 
= e^{2i\stackrel{(1)}{\delta }}=1+2i\stackrel{(1)}{a},
\label{eq:sdel}
\end{eqnarray}
by the BW formula
\footnote{
Here we apply a relativistic form with a modification, taking into 
consideration the indefiniteness of the masses of unstable particles. 
Decay widths of resonances will be estimated using this form in \S 3.
}
\begin{eqnarray}
\stackrel{(1)}{a} &=& \frac{
-\sqrt{s} \Gamma_{(1)}^1 (s)}
{s-M_{(1)}^2+i\sqrt{s} \Gamma_{(1)}^1 (s)},\ \ \   
\sqrt{s} \Gamma_{(1)}^1 (s)=
\frac{|{\bf p}_1|}{8\pi \sqrt{s}} \stackrel{(1)}{g}^2_1
=\rho_1 (s)\stackrel{(1)}{g}^2_1.
\label{eq:bw}
\end{eqnarray}

The $\stackrel{(1)}{\delta}$ is real, as is seen by substituting 
Eq.(\ref{eq:bw}) into Eq.(\ref{eq:sdel}),
and the unitarity relation $SS^*=1$, Eq.({\ref{eq:unisl}), is easily 
shown to be satisfied.

In the two-resonance case the scattering matrix,
\begin{eqnarray}
S &=& e^{2i\delta (s)}=1+2ia(s) ,
\label{eq:sarel}
\end{eqnarray}
is given as a multiplicative
form of the ``respective resonance $S$-matrices" 
$\stackrel{(i)}{S}=e^{2i\stackrel{(i)}{\delta}}=1+2i\stackrel{(i)}{a}$ as
\begin{eqnarray}
S &=&\stackrel{(1)}{S}\stackrel{(2)}{S}.
\label{eq:smul}
\end{eqnarray}
The unitarity relation $SS^* =1$
is easily seen to be satisfied by
the ``unitarity of respective $\stackrel{(i)}{S}$."
The scattering amplitude $a(s)$ 
($=-\rho_1 (s) {\cal T}(s)$) is represented, in terms of the respective
BW resonances, as
\begin{eqnarray}
a(s)=
\stackrel{(1)}{a}(s)+ 
\stackrel{(2)}{a}(s)+
2i 
\stackrel{(1)}{a}(s) 
\stackrel{(2)}{a}(s).
\label{eq:2resa}
\end{eqnarray}
This amplitude consists of
a simple sum of respective BW resonances and
their cross term ($ 2i \stackrel{(1)}{a}(s) \stackrel{(2)}{a}(s)$).
The latter, which looks like
an ``interference" of the former two resonances, appears
in the amplitude. Thus we call this method of
constructing the amplitudes the Interfering-Amplitude (IA)
method. 

Extension to the three-resonance case is straightforward, and the 
formulae are given as
\begin{eqnarray}
S&=& e^{2i(\stackrel{(1)}{\delta}+
\stackrel{(2)}{\delta}+\stackrel{(3)}{\delta})}
=\stackrel{(1)}{S}\stackrel{(2)}{S}\stackrel{(3)}{S},\\
a(s)&=&
\stackrel{(1)}{a}   + 
\stackrel{(2)}{a}   +
\stackrel{(3)}{a}   +
2i(
\stackrel{(1)}{a}    
\stackrel{(2)}{a}   +
\stackrel{(2)}{a}    
\stackrel{(3)}{a}   +
\stackrel{(3)}{a}    
\stackrel{(1)}{a}    )
-4
\stackrel{(1)}{a}    
\stackrel{(2)}{a}    
\stackrel{(3)}{a}.   
\label{eq:3resa}
\end{eqnarray}

A background phase $\delta^{BG}$ is added
to the phase shift Eq.(\ref{eq:delta}), if necessary:
\begin{eqnarray}
\delta =\delta^{Re} + \delta^{BG}.
\label{eq:drebg}
\end{eqnarray}
The unitarity relation is satisfied for any real $\delta^{BG}$. 
In this case the scattering amplitude is given by 
a background amplitude $a^{BG}$ 
and the resonance amplitude $a^{Re}$ (Eqs.(\ref{eq:2resa}) and 
(\ref{eq:3resa})) as
\begin{eqnarray}
a = a^{BG} + a^{Re} + 2ia^{BG} a^{Re}= 
a^{BG} + a^{Re} e^{2i\delta^{BG}},
\label{eq:bgp}
\end{eqnarray}
where
\begin{eqnarray*}
a^{BG}\equiv\frac{e^{2i\delta^{BG}}-1}{2i}.
\end{eqnarray*}
Application to the greater than three-resonance case can be done
in a similar way.

\subsection{two-channel multi-resonance scattering amplitude}
Next we construct an explicit form of the scattering amplitude 
in the two-channel case. 
We treat scattering matrix elements
$S_{11}$, $S_{12}$ and $S_{22}$.
The element $S_{21}$ is equal to $S_{12}$ by time reversal invariance
(Eq.(\ref{eq:tri})).
On these three elements is imposed the unitarity relation
(Eq.(\ref{eq:unisl}))
\begin{eqnarray}
|S_{11}|^2+
|S_{12}|^2&=&
|S_{22}|^2+
|S_{12}|^2=1, \nonumber \\
S_{11}S_{12}^* &+& S_{12}S_{22}^*=0.
\label{eq:conp}
\end{eqnarray}
The form of $S_{ij}$, satisfying the above equations,
is written as
\begin{eqnarray}
S_{11}&\equiv&\eta e^{2i\delta},\nonumber \\
S_{12}&\equiv&\sqrt{1-\eta^2}e^{i(\phi +\frac{\pi}{2})},
\nonumber \\
S_{22}&\equiv& \eta e^{2i\delta '},
\label{eq:elas}
\end{eqnarray}
in terms of the phase shift $\delta (\delta ')$ in the 
$\pi\pi (K\overline{K})$ scattering, the corresponding elasticity
$\eta (=\eta ')$, and the phase $\phi$ 
in the $\pi\pi\rightarrow K\overline{K}$ scattering.
These phases are constrained by
\begin{eqnarray}
\delta + \delta ' = \phi.
\label{eq:dsum}
\end{eqnarray}

In the one-resonance case, the scattering matrix elements satisfying
the unitarity relation Eq.(\ref{eq:conp})
are given in terms of the BW amplitude as
\begin{eqnarray}
\stackrel{(1)}{S}_{11}&=&1+2i\stackrel{(1)}{a}_{11},\ 
\stackrel{(1)}{S}_{12}=2i\stackrel{(1)}
{a}_{12},\ 
\stackrel{(1)}{S}_{22}=1+2i\stackrel{(1)}{a}_{22},
\label{eq:sadef}\\
\stackrel{(1)}{a}_{ij}&=&
\frac{
-\sqrt{
\sqrt{s}\Gamma_{(1)}^i
\sqrt{s}\Gamma_{(1)}^j}
}
{s-M_{(1)}^2+i\sqrt{s}\Gamma_{(1)}^{tot} (s)},\ \ \ 
\Gamma_{(1)}^{tot}(s)=\Gamma_{(1)}^1(s)+\Gamma_{(1)}^2(s).
\label{eq:adef}
\end{eqnarray}
Phases $\delta$ and $\phi$ are obtained by comparing 
Eqs.(\ref{eq:sadef}) and (\ref{eq:adef}) with their definition 
Eq.(\ref{eq:elas}), and
the elasticity $\eta$ 
is given as
\begin{eqnarray}
\stackrel{(1)}{\eta}=
\sqrt{1-4\frac{
\sqrt{s}\Gamma_{(1)}^1 (s) \sqrt{s}\Gamma_{(1)}^2 (s) }
{(s-M_{(1)}^2)^2 + (\sqrt{s}\Gamma_{(1)}^{tot}(s))^2} }.
\end{eqnarray}

In the two-resonance case, extending the method in Eq.(\ref{eq:smul})
to this case, the diagonal elements of the scattering matrix,
$S_{11}$ and $S_{22}$, are represented in a multiplicative form of 
respective resonance $S$-matrices
\begin{subeqnarray}
\label{eq:srep}
S_{11} &=& \stackrel{(1)}{S}_{11}\stackrel{(2)}{S}_{11},\ 
S_{22} = \stackrel{(1)}{S}_{22}\stackrel{(2)}{S}_{22}, \\
\stackrel{(i)}{S}_{11} 
&=& \stackrel{(i)}{\eta} e^{2i\stackrel{(i)}{\delta}},\ 
\stackrel{(i)}{S}_{22} 
    =\stackrel{(i)}{\eta} e^{2i\stackrel{(i)}{\delta '}},
\end{subeqnarray}
where the elasticity $\eta$ in each diagonal channel
is represented in the multiplicative form 
$\eta=
\stackrel{(1)}{\eta}
\stackrel{(2)}{\eta}$, and the phase shift as a sum 
$\delta =\stackrel{(1)}{\delta}+\stackrel{(2)}{\delta}$, following
from its physical meaning. 
Corresponding amplitudes are given in a form similar to 
Eq.(\ref{eq:2resa}) in the one-channel case:
\begin{eqnarray}
a_{11}&=&\frac{S_{11}-1}{2i}=
\stackrel{(1)}{a}_{11}+ 
\stackrel{(2)}{a}_{11}+
2i 
\stackrel{(1)}{a}_{11} 
\stackrel{(2)}{a}_{11},\nonumber \\
a_{22}&=&\frac{S_{22}-1}{2i}=
\stackrel{(1)}{a}_{22}+ 
\stackrel{(2)}{a}_{22}+
2i 
\stackrel{(1)}{a}_{22} 
\stackrel{(2)}{a}_{22}.
\end{eqnarray}
The non-diagonal $S$-matrix element is determined through relations
(\ref{eq:elas}) and (\ref{eq:dsum}) with the diagonal elements given above.
First, the phase shift $\phi$ of $S_{12}$ is
equal to a sum of respective contributions of resonances:
$\phi=
\stackrel{(1)}{\phi}+
\stackrel{(2)}{\phi}$.
So it becomes of the form $S_{12}=2i\stackrel{(1)}{a}_{12} \times 
\stackrel{(2)}{a}_{12}
\times$(real function).
The absolute value of $S_{12}$ is determined by Eq.(\ref{eq:conp})
\begin{eqnarray}
S_{12}=
-i
\stackrel{(1)}{a}_{12}
\stackrel{(2)}{a}_{12}
\frac{\sqrt{1-\eta^2}}
{|
\stackrel{(1)}{a}_{12}
\stackrel{(2)}{a}_{12}
|}
\equiv 2ia_{12}.
\label{eq:sdef}
\end{eqnarray}
The unitarity Eq.(\ref{eq:conp}) is guaranteed by the above procedure
to derive $S_{12}$.

Extension to the three-resonance case can be done in the same manner.
The results are given as
\begin{eqnarray}
\label{eq:3ras}
S_{ii}&=&
\stackrel{(1)}{S}_{ii}
\stackrel{(2)}{S}_{ii}
\stackrel{(3)}{S}_{ii}\ \ \ {\rm for}\ \ i=1,2\ , \nonumber \\
S_{12}&=&
-i
\stackrel{(1)}{a}_{12}
\stackrel{(2)}{a}_{12}
\stackrel{(3)}{a}_{12}
\frac{\sqrt{1-\eta^2}}{|
\stackrel{(1)}{a}_{12}
\stackrel{(2)}{a}_{12}
\stackrel{(3)}{a}_{12}
|}, \nonumber \\
\eta&=&
\stackrel{(1)}{\eta}
\stackrel{(2)}{\eta}
\stackrel{(3)}{\eta},
\end{eqnarray}
\begin{eqnarray}
\label{eq:3ras2}
a_{ii}&=&
\stackrel{(1)}{a}_{ii}   + 
\stackrel{(2)}{a}_{ii}   +
\stackrel{(3)}{a}_{ii}   +
2i(
\stackrel{(1)}{a}_{ii}    
\stackrel{(2)}{a}_{ii}   +
\stackrel{(2)}{a}_{ii}    
\stackrel{(3)}{a}_{ii}   +
\stackrel{(3)}{a}_{ii}    
\stackrel{(1)}{a}_{ii}    )
-4
\stackrel{(1)}{a}_{ii}    
\stackrel{(2)}{a}_{ii}    
\stackrel{(3)}{a}_{ii}, \nonumber \\
a_{12}&=&\frac{S_{12}}{2i}.
\end{eqnarray}
Here, it may be worthwhile to note that all phases, $\delta (s),\delta '(s)$
and $\phi (s)$, are given as a sum of contributions of respective 
resonances as in Eq.(\ref{eq:delta}).
Application to the greater than three-resonance case
can be done in a similar way.

Background phases are added 
to phase shifts if they are necessary. In the two-channel case,
in order to satisfy
time reversal invariance, that is $S_{12}=S_{21}$, 
background phases are introduced as
\begin{eqnarray}
S&=&S^{BG}S^{Re}S^{BG},\nonumber \\
S^{BG}&=&
\left(
\begin{array}{cc}
e^{i\delta^{BG}} & 0 \\
0 & e^{i\delta'^{BG}}
\end{array}
\right) ,
\end{eqnarray}
where $\delta^{BG}$ ($\delta '^{BG}$)
denotes a phase in the $\pi\pi$ ($K\overline{K}$) channel.
This leads to the following change of the $S$-matrix phase:
\begin{eqnarray}
S_{11}^{Re}&\rightarrow&S_{11}^{Re}e^{2i\delta^{BG}}\nonumber \\
S_{12}^{Re}&\rightarrow&S_{12}^{Re}e^{i(\delta^{BG}+\delta '^{BG})}\nonumber \\
S_{22}^{Re}&\rightarrow&S_{22}^{Re}e^{2i\delta '^{BG}}.
\end{eqnarray}
In the actual analysis we introduce a background phase only in the $\pi\pi$
channel, that is, $\delta '^{BG}=0$. In this case we obtain
\begin{eqnarray}
\label{eq:bgp2c}
a_{11}&=& a_{11}^{BG} + a_{11}^{Re} e^{2i\delta^{BG}}, \nonumber \\
a_{12}&=& a_{12}^{Re} e^{i\delta^{BG}}, \nonumber \\
a_{22}&=& a_{22}^{Re},
\end{eqnarray}
where $a_{11}^{BG}=\frac{e^{2i\delta^{BG}} -1}{2i}$.

Finally, in this subsection we rewrite our formulae of
the two-channel scattering matrix in a simpler form in terms of
\begin{eqnarray}
\stackrel{(i)}{d} &=&
s-M_{(i)}^2+i\sqrt{s}\Gamma_{(i)}^{tot} (s),\nonumber \\
\stackrel{(i)}{n} &=&
s-M_{(i)}^2+i
(-\sqrt{s}\Gamma_{(i)}^1 (s)+
\sqrt{s}\Gamma_{(i)}^2 (s)).
\label{eq:nddef}
\end{eqnarray}
In the one-resonance case they are given as
\begin{eqnarray}
S_{11} = 
\stackrel{(1)}{n}
/
\stackrel{(1)}{d}, \ 
S_{22} =
\stackrel{(1)}{n}^* /
\stackrel{(1)}{d}, \
S_{12} = -i
\sqrt{
|\stackrel{(1)}{d}|^2-
|\stackrel{(1)}{n}|^2}\ /
\stackrel{(1)}{d}.
\label{eq:snd}
\end{eqnarray}
In the multi-resonance case they are
\begin{eqnarray}
S_{11} &=&\prod_i 
\stackrel{(i)}{n}
/
\stackrel{(i)}{d}, \ 
S_{22} =\prod_i 
\stackrel{(i)}{n}^*/
\stackrel{(i)}{d}, \nonumber \\
S_{12} &=& -i
\sqrt{\prod_i
|\stackrel{(i)}{d}|^2-
\prod_i |\stackrel{(i)}{n}|^2}\ /
\prod_i \stackrel{(i)}{d}.
\label{eq:prod}
\end{eqnarray}
The above form of the one resonance scattering matrix 
coincides with the one given in 
Ref. \citen{rf:ff}, whose method can be naturally extended to
the multi-resonance case and gives the same form as Eq.(\ref{eq:prod}).

\section{Analysis of $\pi\pi$-scattering phase shift}

Figure \ref{fig:cm} shows the $S$-wave, I=0 $\pi\pi$ phase shift
reported in the analysis of the CERN-M\"unich data,\cite{rf:CM}
which is widely accepted as a standard of the phase shift up to 
1900MeV.

It shows a rapid step up by $180^\circ$
near the $K\overline{K}$ threshold,
following a slow increasing
which reaches $90^\circ$ slightly below 900MeV. 
Above the $K\overline{K}$ threshold,
it continues 
\begin{wrapfigure}{r}{6.6cm}
  \epsfysize=3.3in
 \centerline{\epsffile{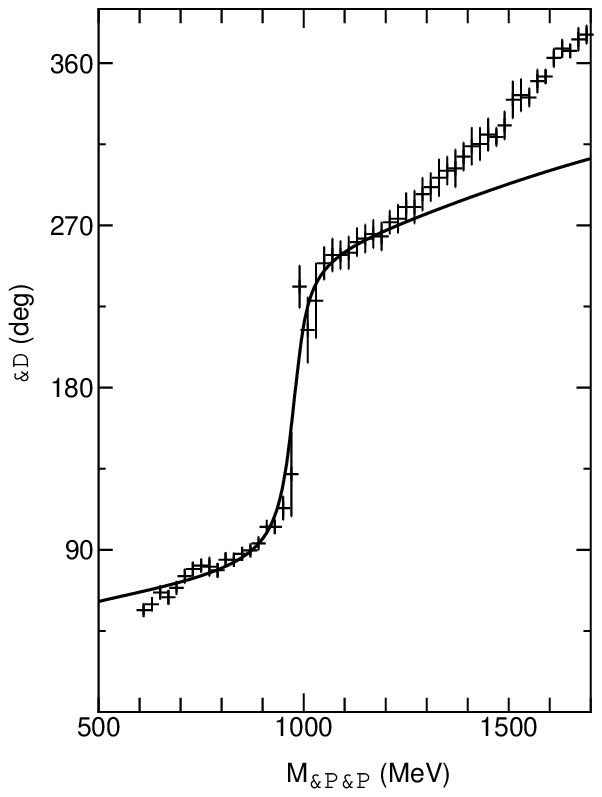}}
 \caption{$\pi\pi$ scattering phase shift. The fitting by the conventional 
interpretation with
$f_0(980)$ and $f_0(1300)$ is also shown.}
 \label{fig:cm}
\end{wrapfigure}
\noindent to grow slowly 
up to about $270^\circ$ at 1300MeV and $360^\circ$ at 1600MeV. 
The increase in the phase shift by
$180^\circ$ corresponds to one full-resonance.
\footnote{We assume here that our relevant resonance particles couple with the
$\pi\pi$ channel strongly enough to make this phase shift.}
Accordingly,
the phase shift up to 1300MeV ($\sim 270^\circ$) has been interpreted
so far as owing to the existence of two resonances, {\em i.e.}
a narrow $f_0(980)$ (full-contribution by $180^\circ$ because of its narrowness) 
and a very broad $f_0(1300)$ (half-contribution by $90^\circ$).  
The latter, which corresponds to $\epsilon(1200)$ in 1976 PDG\cite{rf:pdg76},
replaced the pole previously listed below 600MeV, and $\sigma$
has been omitted from the PDG table since the early 1970's.

However, we have found some
difficulties in the conventional interpretation mentioned above. 
We have made a fitting of phase shift below 1300MeV following this 
interpretation, for which the results are given in Fig.\ref{fig:cm}.
As is clearly seen, the significant structure of
a small bump of the phase shift around 750MeV
cannot be reproduced
by this interpretation.
This structure clearly requires certain sources around
this mass region.
Moreover, it seems that the behavior of phase shift
below 600MeV, which will be given in \S 3.1,
is not reproduced by this fitting.

To solve these two problems, in analogy with the case of nuclear forces,
we have introduced
a negative background source, which works to pull back the phase shift,
\begin{eqnarray}
\delta^{BG}=-r_c|{\bf P}_{\pi\pi}|,
\label{eq:bgdef}
\end{eqnarray}
where $r_c$ is a constant parameter with dimension of length,
and ${\bf P}_{\pi\pi}$ is equivalent to ${\bf p}_1$ in Eq.({\ref{eq:rhom}).
Then the phase and amplitude are described, including 
the background, as in Eq.(\ref{eq:drebg})
and Eqs.(\ref{eq:bgp}) and (\ref{eq:bgp2c}), respectively.
The source of this negative phase background must 
be a ``repulsive" force between 
pions. We will give some arguments on its possible
physical origin in \S 4(C).

In the following, we perform fittings on the various data
using the IA-method with 
the background phase. In \S 3.1, an analysis of the $\pi\pi$
phase shift 
under the $K\overline{K}$ threshold is presented, focusing on the possible 
existence of $\sigma(555)$.
In \S 3.2, the 
data including those over the $K\overline{K}$ threshold, the $\pi\pi$ 
phase shift, the elasticity and the $\pi\pi\rightarrow K\overline{K}$
phase shift, are examined all together
in order to investigate the properties of $f_0(980)$.
Finally, in \S 3.3, a more detailed study is made on the shape
of the negative background, trying a
fitting in the entire relevant mass region.

\subsection{Fitting $\delta$ below the $K\overline{K}$ threshold
---Existence of the $\sigma(555)$ particle and its properties}
\label{sec:und}

\begin{wrapfigure}{r}{6.6cm}
  \epsfysize=9.0 cm
 \centerline{\epsffile{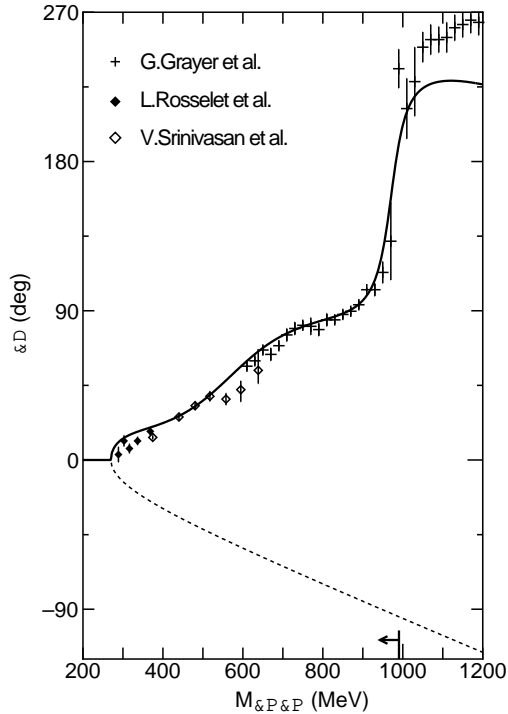}}
 \caption{Fitting of $\delta$ below the $K\overline{K}$ threshold
  (shown by an arrow)
  with two resonances. Behavior of the negative background phase
is also shown (dashed line). The parameters used are described in
the main text. Their best fitted values are listed in Table I.}
 \label{fig:2bwd}
\end{wrapfigure}
We first perform a fitting for the $\pi\pi\rightarrow\pi\pi$ 
phase shift ($\delta$) below the $K\overline{K}$ threshold
(280MeV--980MeV).
In this region, we suppose the existence of 
two resonances, $\sigma$ and $f_0(980)$, and introduce the negative
background mentioned above. 

We apply the one-channel two-resonance 
formula\footnote{See the remarks to be given in \S \ref{sec:anc}}
given in 
Eqs.(\ref{eq:bw}), (\ref{eq:sarel}) and (\ref{eq:2resa}) with the background
phase described in Eqs.(\ref{eq:bgp}) and (\ref{eq:bgdef}).
Necessary parameters to be determined are the following five:
$M_c(=M_{\scriptscriptstyle (1)})$, 
$g_{c\pi\pi}(=\stackrel{\scriptscriptstyle (1)}{g}_1)$, 
$M_g(=M_{\scriptscriptstyle (2)})$, 
$g_{g\pi\pi}(=\stackrel{\scriptscriptstyle (2)}{g}_1)$ and $r_c$;
$g_{\pi\pi}$ is the coupling constant to $\pi\pi$ channel, where
suffix $c$ and $g$ represent $\sigma$ and $f_0(980)$, respectively. 
$r_c$ is the ``radius" of the negative background.
Coupling to $K\overline{K}$
is not needed, because we only treat the phase shift below the threshold
here.

The CERN-M\"unich analysis provides data in the mass region over 600MeV.
For acquiring the property of $\sigma$,
we supplement the data from the $\pi\pi$ threshold (280MeV) 
to 600MeV by Srinivasan 
{\em et al.}\cite{rf:srini} and by Rosselet {\em et al.}\cite{rf:rossel}

Figure \ref{fig:2bwd} shows the result of our fitting. Values of
resonance parameters obtained are 
listed in Table I.
The most notable result is identification of $\sigma$ with 
the low mass of $553.3\pm 0.5_{st}{\rm MeV}$.
\footnote{There are several choices of the relevant
relativistic Breit Wigner forms.
The mass value is dependent to some extent upon these choices.} 
According to the result of another fitting to data including those 
over the $K\overline{K}$ threshold, which will be presented 
in the next sub-section, the
$K\overline{K}$-coupling of $\sigma(555)$ turned out to be 
negligible. Thus the total width 
$242.6\pm 1.2_{st} {\rm MeV}$ owes only to its
$\pi\pi$ coupling.
The fine structure of the phase shift
around $280\sim 900$MeV, especially the bump
around 750MeV, is reproduced quite well.  
It is due to the cooperation of the phase increase by $\sigma (555)$
and decrease by the negative background.
The radius of the negative background turned out to be $3.46{\rm GeV}^{-1}$.

\begin{table}[t]
\begin{center}
\caption{Parameters in the best fit below the $K\overline{K}$ threshold with 
2 resonances and the negative background phase.
Properties of $f_0(980)$ in this table may not be definitive, 
due to omitting data over the $K\overline{K}$ threshold.
For the property of $f_0(980)$, see \S 3.2 and Table II.
The value of $g_{K\overline{K}}$ is quoted from the result of fitting
data including those over the $K\overline{K}$ threshold.
The decay width $\Gamma$ and the ``peak width'' $\Gamma^{(p)}$ are
defined, respectively, as 
$\Gamma_i=\int_{0}^{\infty} ds
\Gamma_i(s)s^{1/2}\Gamma_i^{tot}(s) / (\pi
[(s-M_i^2)^2+s\Gamma_i^{tot}(s)^2])$ 
and $\Gamma_i^{(p)}=\Gamma_i(s=M_i^2)$.
The relation of g and $\Gamma(s)$ is given in our BW formula 
Eq.(2.14).
If we use another form of BW formula, instead of 
(2.14),
as $-M\Gamma(M^2) / (s-M^2+iM\Gamma(M^2))$, the pole position is simply
given as $s_{pole}/M$=$M-i\Gamma(M^2)$. See the discussions in \S 4(A).
}
 \begin{tabular}{|l|c|c|c|c|c|} \hline
          & mass(MeV)  & $g_{\pi\pi}$(MeV)
          & $g_{K\overline{K}}/g_{\pi\pi}$ 
          & $\Gamma_{tot}({\rm MeV})$=$\Gamma_{\pi\pi}$  
          & $\Gamma^{(p)}({\rm MeV})$\\ 
 \hline
 $\sigma$ & 553.3$\pm$0.5 & 3336$\pm$12 & 0.& 242.6$\pm$1.2 
 & 349.3$\pm$2.5 \\
 \hline
 \cline{4-6}
 $(f_0(980))$ & (970.7$\pm$2.2) &\multicolumn{1}{c|}{(1768$\pm$24)}  \\ 
 \cline{1-3}
 \end{tabular}
 \begin{tabular}{|c|c|}
 \hline
    & 2 BW(${\rm GeV}^{-1}$) \\
 \hline
 $r_c$  & 3.46$\pm$0.01  \\
 \hline
 \end{tabular}
 \end{center}
 \label{tab:2bwd}
\end{table}

\subsection{Fitting of $\delta$, $\eta$ and $\phi$ over the $K\overline{K}$
threshold
---properties of $f_0(980)$}
\label{sec:over}

In the following we will investigate the
$\pi\pi\rightarrow\pi\pi$ phase shift ($\delta$), the
elasticity ($\eta$), and the
$\pi\pi\rightarrow K\overline{K}$ phase shift 
($\phi$) 
over the $K\overline{K}$ threshold
all together,
to study the properties
of $f_0(980)$.
We shall try to fit data in the mass region from 600MeV to 1300MeV
\footnote{See the caption of Fig.\ref{fig:3bw}.}
with three resonances: $\sigma$, $f_0(980)$ and $f_0(1300)$.
We apply the two-channel
\footnote{
Below 1100MeV, the intermediate states in the final state interaction
may be limited to the $\pi\pi$ and
$K\overline{K}$ channels. We will treat only these two channels up to 1300MeV.}
three-resonance formula
\footnote{See the discussion to be given in \S \ref{sec:anc}}
given as Eqs.
(\ref{eq:3ras}) and (\ref{eq:3ras2}) with a background phase (\ref{eq:bgp2c})
and (\ref{eq:bgdef}).
Thus the total number of parameters is 10: 
$M_c(=M_{\scriptscriptstyle (1)})$, 
$M_g(=M_{\scriptscriptstyle (2)})$, 
$M_0(=M_{\scriptscriptstyle (3)})$, 
$g_{c\pi\pi}(=\stackrel{\scriptscriptstyle (1)}{g}_1)$, 
$g_{g\pi\pi}(=\stackrel{\scriptscriptstyle (2)}{g}_1)$, 
$g_{0\pi\pi}(=\stackrel{\scriptscriptstyle (3)}{g}_1)$, 
$g_{cK\overline{K}}(=\stackrel{\scriptscriptstyle (1)}{g}_2)$,
$g_{gK\overline{K}}(=\stackrel{\scriptscriptstyle (2)}{g}_2)$,
$g_{0K\overline{K}}(=\stackrel{\scriptscriptstyle (3)}{g}_2)$
and $r_c$.\\

\noindent {\underline{\it Data treatment}}

\begin{figure}[t]
  \epsfysize=4.3in
 \centerline{\epsffile{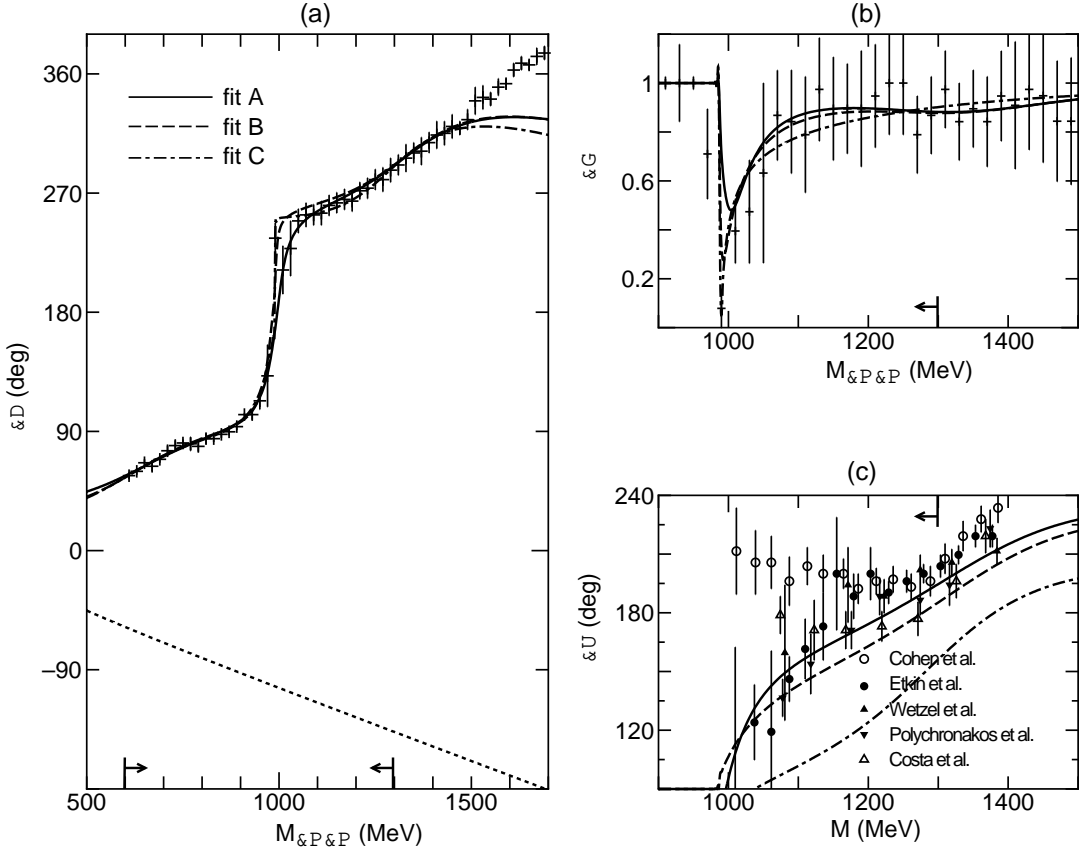}}
 \caption{The fitting for (a) $\delta$, (b) $\eta$, and (c) $\phi$ 
 over the $K\overline{K}$ threshold with three resonances 
(fit A: without the two data points (mentioned in the text), fit B: with
 all data points, fit C: omitting $\phi$ behavior).
 Data in $600{\rm MeV}\sim 1300{\rm MeV}$ shown by arrow 
 are used for the fitting (the data over 
 1300MeV are omitted 
 because we must consider the possible effect due to the 
 other resonances in that mass region).
 }
 \label{fig:3bw}
\end{figure}
Our treatment of data is as follows:\\
(a) $\delta$ over 600MeV
\footnote{The procedure omitting the data below 600MeV has
some influence on 
the mass and width of $\sigma$.
It implies that the behavior of the background phase is not as simple as 
in Eq.(\ref{eq:bgdef}).
This problem will be discussed in \S \ref{sec:soft} 
Anyway, as far as inquiring into the physical problem 
over and around the $K\overline{K}$ threshold,
the low energy region below 600MeV 
is not so important. Thus we use only
the CERN-M\"unich data for fitting here.}
is quoted from the CERN-M\"unich analysis.\cite{rf:CM} 
As can be seen in Fig.\ref{fig:cm},
one data point just on the cliff of $f_0(980)$ exhibits strange behavior.
It looks like a sharp peak, which might not happen in a physical phase
shift. 
Including or excluding this point seriously influences the
results on $f_0(980)$.
Thus we try the following two types of fitting, {\em i.e.}, one which does not
include
this data point (fit A) and another which uses all data points
(fit B).
\\
(b) $\eta$ is quoted from the same analysis. Also there is a point 
corresponding to the point referred in (a) just 
in the narrow dip about 1GeV, which seems to require quite a large inelasticity. 
In fit A we will not use this point and in fit B we will use it.\\
(c) $\phi$ is quoted from Ref.\citen{rf:morg}.
As can be seen in Fig.\ref{fig:3bw}(c), different conflicting results are
reported on the $\phi$-behavior. Thus
we shall perform a compromised fitting: we
use a simple mean of the $\phi$ values 
by Etkin {\em et al.}\cite{rf:etkin}(${\cal E}$)
and by Cohen {\em et al.}\cite{rf:cohen}(${\cal C}$), a square-mean 
of their errors as statistical errors, and their differences
as systematic errors 
: $mean =({\cal E}+{\cal C})/2 $, 
$error = \sqrt{ (\Delta {\cal E})^2 +(\Delta {\cal C})^2 }
+|{\cal E}-{\cal C}|$.\\
As will be shown later, our result reproduces that of
Etkin {\em et al.} rather than that of Cohen {\em et al.} 
The fitting by Morgan and Pennington\cite{rf:morg}
also exhibits a similar tendency.
Our result does not change significantly when 
using only the data of Etkin {\em et al.} \\

\noindent {\underline{\it Phase $\delta$ supplied from negative background}}

Fitting of $\delta$ is shown in Fig.\ref{fig:3bw}(a).
Concerning our relevant problem, there is no significant difference
between the cases in which we use fit A and B.

Note that
we have enough space of $\delta$ 
for 3 resonances: $\sigma$, $f_0(980)$ and $f_0(1300)$
below 1300MeV, due to the
contribution of the negative background.  
The other resonances,
\footnote{However, they may be 
``partial resonances", which couple with
the $\pi\pi$ states only partially,
when the corresponding phase goes up and then down, and finally
giving no contribution to the phase shift.} 
such as a glueball candidate
can obviously exist with the mass 
around $1500\sim1600{\rm MeV}$.\\

\begin{table}[t]
 \begin{center}
 \caption{Parameters in the best fit over the $K\overline{K}$ threshold
with 3 resonances and the negative background phase.
The results on $\sigma$ in this table are somewhat different from
those in \S 3.1, due to omitting the data below 600MeV. This problem is 
solved in \S 3.3. For reliable results on $\sigma$, see Table I.\ \  
The results on $f_0(1300)$ may not be definitive due to the uncertainty
in the choice of background phase. See the main text.\ \ 
On the  definition of $\Gamma$ and $\Gamma^{(p)}$, see the caption in 
Table I.
}
 \begin{tabular}{|lc|c|cc|c|cc|} \hline
         & & mass(MeV)  & $g_{\pi\pi}$(MeV) & $g_{K\overline{K}}/g_{\pi\pi}$\\ 
 \hline
 $(\sigma)$ & &(598.6$\pm$15.6-29.6) & (4126$\pm$348) & (0.0$\pm$0.349)\\ 
 \hline
$f_0(980)$&  fit A    & 993.2$\pm$6.5$\pm$6.9 & 1680$\pm$91 & 1.48$\pm$0.23 \\
 & fit B    & 982.3$\pm$2.2$\pm$2.4 & 1442$\pm$82 & 2.82$\pm$0.22 \\
 & fit C    & 987 & 1947 & 3.77 \\
 \hline
 $(f_0(1300))$ & & (1313$\pm$36) & (5185$\pm$272) & (0.286$\pm$0.079)\\
 \hline
 \end{tabular}
 \begin{tabular}{|lc|c|cc|} \hline
     & &$\Gamma_{tot}({\rm MeV})$ & $\Gamma_{\pi\pi}$ 
          & $\Gamma_{K\overline{K}}$ \\ 
 \hline
$(\sigma)$ & &  (325.1$\pm$36.3) & (325.1$\pm$36.0) & (0.0$\pm$2.8) \\
 \hline
$f_0(980)$&  fit A &  78.3$\pm$11.5 & 55.70$\pm$6.1 & 22.7$\pm$9.6 \\
 & fit B    & 114.5$\pm$15.5 & 46.1$\pm$4.9 & 68.4$\pm$14.6 \\
 & fit C    &  344 & 73 & 272 \\
 \hline
 $(f_0(1300))$ & &  (372.0$\pm$34.5) & (355.9$\pm$34.2) & (16.1$\pm$9.9) \\
 \hline
 \end{tabular}
 \begin{tabular}{|lc|c|cc|} \hline
     & &$\Gamma_{tot}^{(p)}({\rm MeV})$ & $\Gamma_{\pi\pi}^{(p)}$ 
          & $\Gamma_{K\overline{K}}^{(p)}$ \\ 
 \hline
$(\sigma)$ & &  (505.0$\pm$89.3) & (505.0$\pm$89.3) & (-) \\
 \hline
$f_0(980)$&  fit A &  67.9$\pm$9.4 & 54.4$\pm$6.1 & 13.5$\pm$7.4 \\
 & fit B    & 40.5$\pm$4.7 & 40.5$\pm$4.7 & - \\
 & fit C    &  73.5 & 73.5 & - \\
 \hline
 $(f_0(1300))$ & &  (420.6$\pm$46.2) & (398.6$\pm$44.1) & (22.0$\pm$13.8) \\
 \hline
 \end{tabular}
 \begin{tabular}{|c|c|}
 \hline
    & 3 BW(${\rm GeV}^{-1}$) \\
 \hline
 $r_c$  & 3.76$\pm$0.27  \\
 \hline
 \end{tabular}
 \end{center}
 \label{tab:hcp}
\end{table}

\noindent {\underline{\it Results on $f_0(980)$}}

The mass of $f_0(980)$ is obtained as
$993.2\pm 6.5_{st} \pm 6.9_{sys}$MeV(fit A), where the systematic error 
is estimated as a probable uncertainty ($2\sigma$) of 
$r_c$ (see \S 3.3). 

On the other hand, its coupling to $K\overline{K}$
channel varies a great deal, depending on whether we adopt fit A or B
(Table II).

In Fig.\ref{fig:3bw}(a) the $\delta$ fitting for fit B is also shown.
As can be seen, the fitting over 1GeV is worse than that for fit A, 
due to the influence of the data point 
on the cliff, referred to above.

The elasticity plots for fit A and B are shown in 
Fig.\ref{fig:3bw}(b).
If we take the point on the cliff into account (fit B),
the fitted curve of the elasticity passes through the lower region,
and the $\phi$-fitting becomes slightly worse (Fig.\ref{fig:3bw}(c)). Thus 
the $K\overline{K}$ width and the total width in fit B 
must be larger than those in case of fit A, respectively.
If we intend to reproduce that point exactly,
the $K\overline{K}$ width and total width become
272MeV and 344MeV, respectively (fit C). 
They seem to be consistent with 
recent values obtained by Zou {\em et al.}\cite{rf:zou}
But in this case
it is difficult to reproduce the $\phi$ behavior as shown in the figure.
The obtained width of $f_0(980)$ without this point (fit A) 
is $78.3\pm 11.5_{st}$MeV, which is 
consistent with a value given by
Morgan and Pennington.
\cite{rf:morg}

It should be noted that the
results on properties of $f_0(1300)$ in Table II
are not definitive,
because they
strongly depend on how to choose the background phase,
and also on the properties of other possible $f_0$ with higher masses.

\subsection{Fitting for the entire relevant mass region
--- possible form of negative background}
\label{sec:soft}
As was mentioned in the previous two sub-sections, some differences
of the obtained results
between the fittings in the different mass regions 
suggest that the
shape of background phase is not as simple as Eq.(\ref{eq:bgdef}).

\begin{wrapfigure}{r}{6.6cm}
  \epsfysize=4.3 in
 \centerline{\epsffile{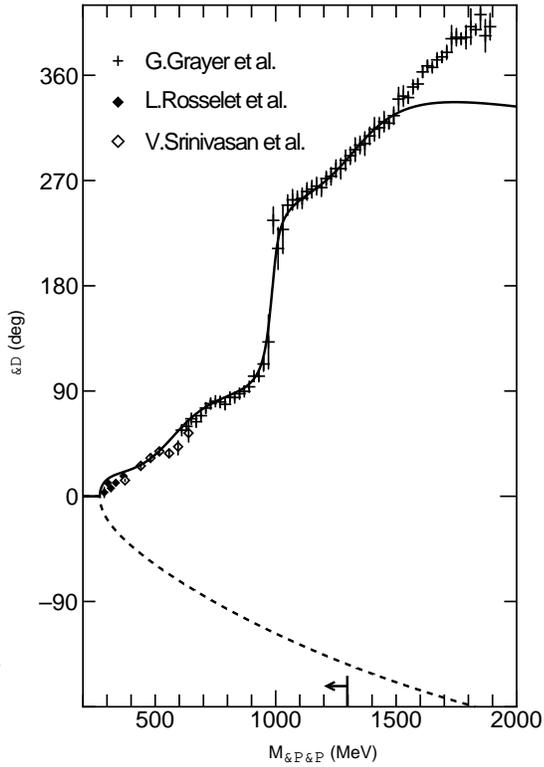}}
 \caption{Fitting for $\delta$ 
 through entire mass region with ``soft" repulsive core. 
 The 
 parameters used 
 for $\sigma$ are quoted from Table I, while those for $f_0(980)$
 and $f_0(1300)$ are from Table II (fit A).}
 \label{fig:soft}
\end{wrapfigure}
From a consideration on the origin of negative background, considering an
analogy with the case of nucleon-nucleon interactions (to be given in \S 4(C)),
we consider it reasonable
that
the radius of the negative background $r_c$ becomes smaller
with the increase of the pion momentum.
Thus we shall further try 
another type of fitting through the whole relevant mass region to determine
a possible shape of the background:\\
(a) We will introduce
an exponentially decreasing factor to 
the radius of the background $r_c$ (Eq.(\ref{eq:bgdef})), 
\begin{eqnarray}
r_c \rightarrow r_c exp(-b|{\bf P}_{\pi\pi}|)\nonumber, \\
\label{eq:soft}
\end{eqnarray}
where the parameter $b$ also has dimensions of length.\\
(b) Parameters giving the properties of three resonances are fixed, 
{\em i.e.},
those for $\sigma$ are taken from the values
obtained from fitting below the $K\overline{K}$ threshold,
while those for $f_0(980)$ and $f_0(1300)$ are from the values
over the threshold.

The result of fitting is shown in Fig.\ref{fig:soft}, where the 
best fit parameters of the background core are given in Table III.
The fitting looks quite good through the entire relevant
mass region. 
\begin{table}[t]
 \begin{center}
 \caption{Parameters in the best fit for the core of Eq.(3.2).
 }
 \begin{tabular}{|c|c|}
 \hline
  $r_c({\rm GeV}^{-1})$  &  $b({\rm GeV}^{-1})$ \\
 \hline
  5.36$\pm$0.01  & 0.480$\pm$0.009  \\
 \hline
 \end{tabular}
 \end{center}
 \label{tab:soft}
\end{table}

\subsection{Remarks on the method of our analysis}
\label{sec:anc}
In performing the analysis in this section we have actually applied
an analytically discontinued (A.D.) density matrix, 
whose elements have non-zero values only
over the threshold of the corresponding channel as
\begin{eqnarray}
\rho&=&diag(
\frac{|{\bf p}_1|}{8\pi\sqrt{s}}, 
\frac{|{\bf p}_2|}{8\pi\sqrt{s}}),\nonumber \\ 
|{\bf p}_1 |&=&\sqrt{\frac{s}{4} -m_\pi^2}\theta(s-4m_\pi^2 ),\  
|{\bf p}_2 |=\sqrt{\frac{s}{4} -m_K^2}\theta(s-4m_K^2 )
\label{eq:rhoad}
\end{eqnarray}
(correspondingly in fitting of \S 3.1 below the 
$K\overline{K}$ threshold we have applied the one-channel formula,
and in fitting of \S 3.2 applied the two-channel formula, since
only the $|{\bf p}_1|$ remains non-zero for the region below 
the $K\overline{K}$ threshold).

Thus the resonance parameters obtained in this analysis are 
considered to represent the particle properties
in the zero-th approximation, which
are determined only through the quark-gluon color-confining dynamics, 
omitting the ``residual"
strong interactions among color-singlet hadrons. However, this treatment
violates the 
``sacred" requirement of analyticity of the scattering
amplitude. Applying an analytically continued (A.C.) density matrix 
is considered 
to take into account the effects of
virtual $\pi\pi$ and $K\overline{K}$ channels in extracting 
the resonant-particle properties.

We have estimated these effects, actually applying the A.C. density matrix:
The result of re-analysis of data (including all of
$\delta$, $\eta$, and $\phi$) in the region 
600$\leq$$\sqrt{s}$$\leq$1300MeV produces little change in the properties
of $\sigma(555)$ and $f_0(980)$. This seems to be due to the zero 
coupling of $\sigma$ to the $K\overline{K}$ channel
({\em cf.} Table I) 
and the small width of $f_0$.
Then, in order to estimate the effects on the $\sigma$-properties by 
applying the A.C. density matrix for $f_0$, we have performed
fitting under the $K\overline{K}$ threshold 
by taking the $f_0$ properties as obtained above and by setting
$g_{K\overline{K}}(\sigma)$=0.
The results are
$M_\sigma$=568.9$\pm$0.5MeV and $\Gamma_\sigma$=250.5$\pm$3.7MeV.
The small differences between these values and those in the case with A.D. 
density matrix
(Table I) owe to the A.C. effect on $f_0$, 
which has non-zero coupling to
$K\overline{K}$.

In the analysis of this section we have not taken into account
the Adler-zero condition\cite{rf:adler} for the amplitude: 
The possible effects of this requirement 
on the properties of $\sigma$ and of $f_0$ are considered
to be small, although it may affect the values of parameters
relevant to the energy region near the $\pi\pi$-threshold.

Here we should point out that it is generally very difficult
to take quantitatively the effects of strong interaction into account,
since hadrons are interacting 
``democratically" 
among various
possible channels (remember the long history of strong
interactions before QCD). Also we feel that the 
physical backgrounds of the analyticity of the scattering
amplitude in the level of point-like hadrons
should be re-examined presently,
when the ``global" 
properties of hadrons are determined through the confinement
dynamics of the quark-gluon world and the hadrons have space-time extensions.

\section{Physical implications of our results and 
related problems}
\noindent (A) {\it Relation of the IA method to the other approaches}\\

The $S$-matrix must satisfy the unitarity (Eq.(\ref{eq:unisl})) 
and be symmetric due to 
time reversal invariance (Eq.(\ref{eq:tri})). The corresponding conditions 
(\ref{eq:unit}) and (\ref{eq:trit}) are
imposed on the ${\cal T}$ matrix, and there have appeared various 
methods\cite{rf:badal}
to parametrize the general solutions satisfying these conditions.

In the most conventional 
${\cal K}$ matrix methods\cite{rf:morg,rf:aniso} 
the ${\cal T}$ matrix is first represented by the ${\cal K}$ matrix as
\begin{eqnarray}
{\cal T}={\cal K} \frac{1}{1+i\rho {\cal K}}
\end{eqnarray}
and is regarded as an analytic function on the Riemann sheets
of complex $s$-variable, 
except for the right and left hand singularities and the poles.
Then the ${\cal T}$ matrix 
for any real symmetric ${\cal K}$ matrix
automatically satisfies the unitarity and the time reversal invariance. 

Our relevant problem is to find pole positions
of the ${\cal T}$ matrix, which are related to
the masses and the decay widths of resonant particles.

However, the parameters used in the ${\cal K}$ matrix are not directly related 
to these properties of physical particles, 
and there seem 
to be many unpleasant features:
Even the number of ${\cal T}$ matrix poles is determined
by the choice of the ${\cal K}$ matrix parametrization.
Moreover, the contributions to the  
${\cal T}$ matrix of the resonances and the non-resonant background cannot 
be separated in the simple ${\cal K}$ matrix representation.
In the Dalitz-Tuan representation\cite{rf:dt} of the ${\cal T}$ matrix, which 
is obtained by modifying the ${\cal K}$ matrix representation,
the background and 
resonance contributions are clearly separated, and the
parametrization used in Ref.\citen{rf:zou}
is equivalent to our Eq.(\ref{eq:bgp2c}) 
in the one-resonance case.

On the other hand, in the $\omega$-method which was proposed by
Kato\cite{rf:kato} and by Fujii and Fukugita,\cite{rf:ff} 
the pole
positions of the $S$-matrix are directly parametrized
on complex Riemann sheets.
They use the Le-Counter Newton equation
for the $S$ matrix in two-channel case,
\begin{eqnarray}
S_{ii}&=& {\bf d}^{(i)}(s)/{\bf d}(s)\ \ {\rm for}\ i=1,2, \nonumber \\
S_{11}&S_{22}&-S_{12}^2={\bf d}^{(1,2)}(s)/{\bf d}(s).
\label{eq:ffsd}
\end{eqnarray}
Here ${\bf d}(s)$, defined on the physical Riemann sheet $I$, includes the
right hand singularities, while the ${\bf d}^{(1)}(s)$, ${\bf d}^{(2)}(s)$ and
${\bf d}^{(1,2)}(s)$ are the analytic
continuations of ${\bf d}(s)$ in the sheet II, IV and III, obtained
by the change of signs of momenta 
$p_1\rightarrow -p_1$,
$p_2\rightarrow -p_2$ and
$(p_1,p_2)\rightarrow (-p_1,-p_2)$.
The four Riemann sheets are mapped onto the single $\omega$-plane\cite{rf:kato}
\begin{eqnarray}
\omega=\frac{1}{\sqrt{m_K^2 -m_\pi^2}} (p_1 +p_2).
\end{eqnarray}
The poles of the $S$ matrix correspond to the zeros of ${\bf d}(s)$ 
in the $\omega$-plane.
The Breit Wigner resonance formula is obtained with ${\bf d}(s)$ 
\begin{eqnarray}
{\bf d}(s)=
\frac{m_K^2-m_\pi^2}{\omega^2} (\omega-\omega_r)(\omega+\omega_r^*)
(\omega-\omega_s)(\omega+\omega_s^*),
\end{eqnarray}
where the
four zeros
($\omega_r$, $\omega_s$ and their inversion for the imaginary axis
$-\omega_r^*$, $-\omega_s^*$)
satisfy certain constraints from the unitarity.
Then, ${\bf d}(s)$ becomes equal to 
$\stackrel{\scriptscriptstyle (i)}{d}(s)$ in Eq.(\ref{eq:nddef}),
if we replace $\sqrt{s}\Gamma_i (s)\equiv\sqrt{s}\frac{p_i}{8\pi s}g_i^2$
by $m\Gamma_i'(s)\equiv m\frac{p_i}{8\pi m^2}g_i^2$, 
and the final form of the $S$ matrix in Eq.(\ref{eq:ffsd})
coincides with our Eq.(\ref{eq:snd}).

In our case with $\sqrt{s}\Gamma_i(s)$ the $S$-matrix has six poles.
Since the $S$-matrix in the IA-method is
described by the simple Breit Wigner resonance formulae and 
is parametrized directly in terms of the masses and coupling constants,
it seems to us not necessary to determine the pole positions of $S$-matrix 
which have no direct physical meanings.\\

\noindent (B) {\it Central production of $\sigma(555)$}\\

According to Watson's final state interaction theorem, 
the phase of production amplitude
of all types of $\pi\pi$ production processes is
determined through
the $\pi\pi$-scattering amplitude.
Accordingly, if the existence of $\sigma(555)$ is really observed
in the $\pi\pi$-scattering, the
corresponding signal must also be observed in any production processes.

In the observations of the
$\pi\pi$ production in $pp$ central collision:
$pp\rightarrow p_f(\pi\pi) p_s$
with hundreds of GeV/c momentum incident proton beam,
a huge event concentration
around $m_{\pi\pi}=0.3\sim 1{\rm GeV}$ has been clearly seen.
\cite{c1,c2,c3} In the conventional treatment thus far made, this
was regarded as a simple 
``background".

However, we find\cite{rf:ti} that the $f_0(980)$ peak 
slides down around 900MeV, probably due
to ``destructive" interference
\footnote{In Ref.\citen{c3} an interference of the $f_0(980)$ with
the $S$-wave background was introduced. 
In Ref.\citen{rf:ti} it is first pointed out that this ``interfering 
background" may originate from a physical particle.}
with the concentration.
We have performed a fitting with two Breit-Wigner resonances
($f_0(980)$ and another of lower mass and wide width), and
found that ``destructive" interference occurs with 
a relative phase between the two resonances, which
seems to be consistent with the above theorem.
This fact supports our interpretation of 
this concentration 
being due to production of
$\sigma(555)$.
However, it is to be noted that the mass and the width 
obtained in Ref.\citen{rf:ti} 
may not be definitive, and the above result should be checked further.
An exact analysis on the $pp$ central collision will 
be reported soon.\cite{rf:cent}

Moreover, by comparing the $\pi\pi$-spectra in the $pp$ central collisions and
in the peripheral $\pi^- p$ charge-exchange processes, it is seen
that the production
probability of $\sigma (555)$ in the former
may be larger than that in the latter.
This fact seems to be
explicable by supposing that
$\sigma(555)$ is a member of the new type scalar nonet,
which will be discussed in (D) below.\\

\noindent (C) {\it ``Repulsive core" in $\pi\pi$-interactions}\\

\begin{table}[t]
 \begin{center}
 \caption{The value of hard core radius
 of $\pi\pi$ interactions, in comparison with
 the cases of $\alpha$-$\alpha$ and $N$-$N$ interactions.
 Structural sizes (EM {\em r.m.s} radii)
 of relevant particles are also shown.\ \ \ 
 * value from Table II.\ \ 
 ** value of soft core from Table III.
 A permissible minimum of the $\pi\pi$ core radius turned to be
 0.5fm ($2.5{\rm GeV}^{-1}$) (see the text). 
 }
 \begin{tabular}{|l|c|c|}
 \hline
  system & $r_c$  &  $\sqrt{<r^2>}$ \\
 \hline
  $\alpha-\alpha$  & $2 fm$ & $1.6 {\rm fm}$ \\
 &  $(10{\rm GeV}^{-1})$ & \\
  $N$-$N$ & $0.4\sim0.5 {\rm fm}$ & $0.8 {\rm fm}$ \\
 &  $(2\sim2.5{\rm GeV}^{-1})$ & \\
\hline 
 $\pi-\pi$ & $0.68 {\rm fm}$ & $0.7 {\rm fm}$ \\
 &  $(3.46{\rm GeV}^{-1})^*$ & \\
 &  $1.06 {\rm fm} $ & \\
 &  $(5.36{\rm GeV}^{-1})^{**}$ & \\
 \hline
 \end{tabular}
 \end{center}
 \label{tab:phc}
\end{table}
In our analysis of the $\pi\pi$-scattering phase shift of \S 3, leading
to strong evidence for the existence of $\sigma (555)$, 
introduction of a negative-phase background Eq.(\ref{eq:bgdef}) has
played a crucial role. We consider that a possible
physical origin of this background is of a very
fundamental nature and that a corresponding repulsive force
may be reflecting a
\footnote{This possibility had been noticed in Ref.\citen{rf:166}.}
Fermi-statistics nature of constituent
quarks and anti-quarks of pions in the $\pi\pi$-system:
A repulsive force due to a similar mechanism has been
known since long ago to exist in
nucleus-nucleus interactions.
In the $\alpha$-$\alpha$ system it is phenomenologically known 
that there exists a
strong short-range repulsion
\cite{rf:165,rf:161,rf:162,rf:163}
which gives a negative scattering
phase shift of the same form as Eq.(\ref{eq:bgdef}) with a corresponding 
size of core $r_c^{\alpha\alpha}=2 {\rm fm}$ 
(, while the {\em r.m.s.} radius, a ``structural size", of $\alpha$ is
$\sqrt{<r^2 >_\alpha}=1.6 {\rm fm}$).
The physical origin of this repulsion was explained to be
\cite{rf:165}
due to the Fermi-statistics property of constituent nucleons of
each $\alpha$-nucleus in the $\alpha$-$\alpha$ system.
Then it may be reasonable to imagine,
\cite{rf:163,rf:25,rf:166}
by replacing
the constituent nucleons in the $\alpha$-particle with constituent
quarks in the nucleon, that there may exist a strong
short-range repulsion in nuclear forces
between nucleon-nucleon system. Actually
it has been shown
\cite{rf:1,rf:4,rf:5}
phenomenologically that there surely exists
a strong repulsive core in $^1S_0$ states. Extensive
investigations\cite{rf:suppl} on nuclear forces have 
been done by many physicists, and the size of 
the corresponding
hard core radius was reported to be $r_c^{NN}=0.4\sim 0.5 {\rm fm}$,
while the electromagnetic (EM) {\em r.m.s.} radius of the proton is
$\sqrt{<r^2>_p}=0.8 {\rm fm}$.
A formulation 
\cite{rf:yazaki}
to derive the short range part of the $N$-$N$ interaction
directly from the quark-quark interaction has also been given.

Now it seems to us quite natural that a possible physical origin
of our repulsive core in the $\pi\pi$ iso-singlet $S$-wave state
is
the same mechanism as that in the nuclear force.
Actually an interesting work from this viewpoint has been
done by Ref.\citen{rf:holin}.
The values of our hard core radii in the $\pi\pi$ interaction are
collected in Table IV with EM {\em r.m.s.} radius of the pion, 
in comparison
with the mentioned cases of the $\alpha$-$\alpha$ and the 
$N$-$N$ interactions.

Hard core radii in the respective cases may be directly related with
and may have similar orders of magnitude to the
structural sizes of corresponding composite particles, considering our
relevant mechanism. This expectation seems
to be realized, as is seen in Table IV. 

However, one may feel that $r_c^{\pi\pi}$ seems to be rather large.
We have made some analysis on this point.
To reproduce the characteristic phase shift below $K\overline{K}$
threshold, a certain magnitude of the core radius is required.
A permissible minimum value of the radius may be $2.5{\rm GeV}^{-1}$.

Here it is noted that our analysis with the ``soft-core" background 
Eq.(\ref{eq:soft}) has been stimulated by the corresponding one 
\cite{rf:31}
in the
case of nuclear force.

We have also estimated the $\pi\pi$ scattering length
$a_{len}$=$a_{len}^{\sigma}$-$r_c$ using the properties
of $\sigma$ in Table I and the values of $r_c$ from both in 
Tables I and III. 
(We have used the form of relativistic B.W. formula,
Eq.(\ref{eq:bw}), in order to estimate $a_{len}^{\sigma}$.)
The results are given in Table V in comparison with the 
other estimations, which seem to be mutually consistent. 
However, we note that the values of $a_{len}^{\sigma}$ depend
largely upon the form of the relativistic B.W. formula,
and effects due to various possible effects of hadron interactions
(see \S \ref{sec:anc}).\\

\begin{table}[t]
 \begin{center}
 \caption{Scattering length obtained from our fitting,
 in comparison with other estimations.}
 \begin{tabular}{|l|c|}
 \hline
  Our estimation                        & 0.69fm ($r_c$=0.68fm) \\
 $a_{len}$=$a_{len}^{\sigma}$-$r_c$ & 0.32fm ($r_c$=1.06fm) \\ 
 \hline
  Other estimation & 0.37$\pm$0.07fm \cite{rf:rossel}\cite{rf:nagels}\\
                 & 0.34$\pm$0.13fm \cite{rf:belkov}\\
\hline
 Current Algebra & $a_{len}$=$\frac{7}{32\pi}\frac{m_\pi}{f_\pi^2}$\\
(Non-linear $\sigma$ model) & =0.22fm\cite{rf:wein,rf:lee}\\
\hline
 \end{tabular}
 \end{center}
 \label{tab:scl}
\end{table}

\noindent (D) {\it Possible existence of chiralons;
new scalar and axial-vector meson nonets}\\

Classification of mesons usually follows the $LS$-coupling scheme
of the non-relativistic quark model.
We generalized it covariantly, keeping good $L$ and $S$
quantum numbers, to the boosted $LS$-coupling scheme.\cite{rf:bls}
In this scheme,
the Bargmann-Wigner spinor wave function,
which is a covariant generalization of Pauli spin functions, 
implies that the constituent quark and anti-quark are
in ``parton-like" parallel motion\cite{rf:spin}, {\em i.e.},
they move with the same velocity
as that of the meson.
``Usual" scalar and axial meson nonets 
can be classified in the first $L$-excited
state in this scheme.

On the other hand, in most ENJL models,\cite{rf:enjl} 
as a low energy effective theory of QCD,
there exist four chiral symmetric nonets with $J^{PC}$=
$0^{++}$, $0^{-+}$, $1^{--}$ and $1^{++}$.
In ENJL essentially only local composite quark 
and anti-quark operators are treated, 
thus missing $L$-excited states in principle.
Here it is to be noted that
the constituent quark and antiquark inside of these
$0^{++}$ and $1^{++}$ nonets
are shown mutually in an anti-parallel motion. 
Members of these nonets are
apparently different from the above-mentioned
Bargmann-Wigner particles

Accordingly, it seems to us that, in order to
complete the meson spectroscopy, we must take 
these extra ``ultra-relativistic" $0^{++}$ and $1^{++}$ nonets into account, 
in addition to the ``non-relativistic" ones described by
the $LS$ coupling scheme.
We call these ``chiralons".\cite{rf:chiralon}

It is expected that
production of chiralons will be enhanced in $pp$ central collisions,
where constituent valence quarks and sea-antiquarks
in anti-parallel motion exist.
Thus 
if we assume that $\sigma$
is a member of scalar chiralons,
enhancement of $\sigma$ in central collisions,
mentioned in (B), may be clearly explained. 

The $f_0(1300)$ given in our analysis in \S 3
may be assigned to a member of the
normal $0^{++}$ nonet in the $^3 P_0$ $q\overline{q}$ state.
Also we are examining the possibility of $f_0(980)$ being a
hybrid meson with a massive constituent gluon.\cite{rf:hyb}

Finally we would like to note that one of the authors has examined \cite{rf:zu}
recently the properties of $\sigma(555)$, obtained in this work, from
the viewpoint of the effective chiral Lagrangian of the linear $\sigma$-model,
and shown that $\sigma(555)$ may be identified with the chiral partner
to the Nambu-Goldstone $\pi$-meson.

\section{Summary}
In this paper we have re-analyzed the phase-shift 
between the $\pi\pi$-threshold and 1300${\rm MeV}$ by applying a newly 
developed
IA-method. The results are summarized as follows:\\
(1) The $\pi \pi \rightarrow \pi \pi$ phase shift has been analyzed
in the mass region of the $\pi\pi$- to the $K\overline{K}$- thresholds,
giving strong evidence for 
the existence of $\sigma$, with mass of $553.3\pm 0.5_{st} {\rm MeV}$ 
and width of $242.6 \pm 1.2_{st} {\rm MeV}$.\\
(2) Our best fit value of a radius of the repulsive core
in $\pi\pi$ interactions,
whose introduction is the keystone for leading 
to the existence of $\sigma(555)$, is
$3.5{\rm GeV}^{-1}$ (in case of ``hard core")/$5.4{\rm GeV}^{-1}$ (``soft core").
This value is
almost equal to the structural size of the pion ($3.5{\rm GeV}^{-1}$). 
This situation is quite similar in the case of $N$-$N$ interactions,
giving support to our interpretation of the repulsive core.\\
(3) The properties of $f_0(980)$ have been investigated from the
data including those over the $K\overline{K}$ threshold.
The mass is obtained as $993.2 \pm 6.5_{st} \pm 6.9_{sys} {\rm MeV}$.
Its total width varies due to the uncertainty of $K\overline{K}$
coupling, to be $78.3\pm 11.5_{st} {\rm MeV}$ (fit A),
$114.5\pm 15.5_{st} {\rm MeV}$ (fit B) or $344 {\rm MeV}$ (fit C),
corresponding to the three different treatments of the
elasticity and the $\pi \pi \rightarrow K\overline{K}$ phase shift, 
both of which have large experimental uncertainties.
However, fit C has difficulty to reproduce the experimental 
$\pi\pi\rightarrow K\overline{K}$ phase behavior.

\section*{Acknowledgements}
We would like to thank H.Shimizu,
who attracted our attention to 
the $\sigma$-particle.
We also thank D.Morgan and M.R.Pennington
who gave us useful comments 
in the preliminary stage of this work.
We appreciate K.Yazaki and Y.Fujii for their 
useful comments on some basic problems.
We are grateful to Y.Totsuka for discussions.
One of the authors (M.Y.Ishida) acknowledges K.Fujikawa and K.Higashijima
for instructive discussions.
Finally the first three of the present authors
should like to thank M.Oda, K.Yamada, N.Honzawa, M.Sekiguchi, H.Sawazaki
and H.Wada for their useful comments and continual encouragement.

\end{document}